\documentclass[10pt]{article}

\usepackage{amsmath}
\usepackage{amssymb}

\usepackage{graphicx}

\usepackage{cite}

\usepackage{color} 

\usepackage[labelfont=bf,labelsep=period,justification=raggedright]{caption}

\makeatletter
\renewcommand{\@biblabel}[1]{\quad#1.}
\makeatother

\date{}

\begin{document}

\begin{flushleft}
{\Large
\textbf{The Association of the Moon and the Sun with Large Earthquakes}
}
\\
Lyndie Chiou$^{1}$, 
\\
\bf{1} ResearchPipeline.com, Union City, CA, USA
\\
$\ast$ E-mail: Corresponding lyndie@researchpipeline.com
\end{flushleft}

\section*{Abstract}
The role of the moon in triggering earthquakes has been studied since the early 1900s.
Theory states that as land tides swept by the moon cross fault lines, stress in the Earth's plates intensifies, increasing the likelihood of small earthquakes. 
This paper studied the association of the moon and sun with larger magnitude earthquakes (magnitude 5 and greater) using a  worldwide dataset from the USGS.

Initially, the positions of the moon and sun were considered separately.
The moon showed a reduction of 1.74\% (95\% confidence) in earthquakes when it was 10 hours behind a longitude on earth and a 1.62\% increase when it was 6 hours behind.
The sun revealed even weaker associations (\textless 1\%).
Binning the data in 6 hours quadrants (matching natural tide cycles) reduced the associations further.

However, combinations of moon-sun positions displayed significant associations.
Cycling the moon and sun in all possible quadrant permutations showed a decrease in earthquakes when they were paired together on the East and West horizons of an earthquake longitude (4.57\% and 2.31\% reductions).
When the moon and sun were on opposite sides of a longitude, there was often a small (about 1\%) increase in earthquakes.

Reducing the bin size from 6 hours to 1 hour produced noisy results. 
By examining the outliers in the data, a pattern emerged that was independent of earthquake longitude.
The results showed a significant decrease (3.33\% less than expected) in earthquakes when the sun was located near the moon.
There was an increase (2.23\%) when the moon and sun were on opposite sides of the Earth.

The association with earthquakes independent of terrestrial longitude suggests that the combined moon-sun tidal forces act deep below the Earth's crust where circumferential forces are weaker.


\section*{Introduction}
There are many proposed triggers for earthquakes. 
Examples include underground temperature differences, solar activity and water movements both below and above ground (see, for example,~\cite{subsurfacetempEQs, glacialearthquakes, earthquakepredictionchapter, earthquakeoverviewchapter, GlobalSeismicAndSolar, EarthPlanet8}).
Tides are especially intriguing as a potential trigger because they are induced by the gravitational pull of the moon and the sun which follow calculable orbits.

Tides occur both in water and, to a lesser extent, in land.
Land tides are a measurable effect wherein the Earth's crust rises and falls as a result of the moon's and sun's gravitational pull. 
The effect can be up to 20 cm near the Earth's poles~\cite{NYTimesLandTides}.

Is it possible that some earthquakes could be triggered by land tides?
It seems a reasonable hypothesis that the intersection of land tides with the Earth's fault lines could serve as a trigger for an earthquake.
The reverse has proven true: moonquakes can be triggered by land tides caused by the Earth's gravitational influence on the moon~\cite{NationalGeographicNews}.

However, the notion that the moon can similarly cause earthquakes is controversial as the Earth has 81 times more mass than the moon~\cite{NASAmoonSite} and any influence is assumed to be very weak at best with many concluding the effect is non-existent (for example, see references~\cite{SpringerLinkNote,JournalGeoResearch1}).

Despite the controversy, papers continue to be published on both sides of the debate (illustrated by~\cite{EarthPlanet5, EarthPlanet6, EarthPlanet7, EarthPlanet1, EarthPlanet2, EarthPlanet3, EarthPlanet4}), many with conclusions based on minimally-sized datasets corresponding to specific faults. 
One noticeable exception was reference~\cite{BigAnalysis} which reported an analysis using 442,412 earthquakes from magnitude 2.5 to 9.
The authors concluded that land tides induced by the moon do trigger earthquakes, albeit primarily shallow earthquakes of low magnitude (a 0.5\% to 1.0\% increase over expected). 
The moon's greatest influence occurred when it was overhead, corresponding to a rise in the Earth's crust.
A short-coming of the analysis was that low magnitude data, particularly below magnitude 4, was not uniformly recorded in the data and potentially lead to a bias in the conclusions.

This paper uses a different dataset than the previous study and not only looks at the role of the moon and sun in triggering earthquakes separately, but extends the analysis to examine the role of the sun and moon in conjunction.
The analysis is restricted to larger earthquakes, magnitude 5 and greater, since these magnitudes were accurately and comprehensively recorded across the globe.
Also, larger magnitude earthquakes can exert a high negative economic impact, making the results more interesting outside of the geological community.

\section*{Methods}

\subsection*{The position of the moon and sun as a proxy for land tides}
A suitable technique to deduce the relationship between land tides and earthquakes would require a dataset that allowed users to extract the land tide value for a given latitude and longitude on a particular date and time.
Unfortunately, no such dataset exists.
The ability to calculate land tides relies on a detailed map of subsurface properties such as ocean floor depths and underground plate tectonics~\cite{OceanMotion}.

For this reason, the longitudinal positions of the moon and sun with respect to earthquakes were used as a proxy for the position of land tides. 
This approach is not the most ideal since, as mentioned, the actual height of land tides varies not just according to the position of the moon and sun, but also due to localized geographical properties.
The assumed correlation of the moon and sun versus land tide heights is an important source of error in the upcoming statistical analysis. 

This paper measures the occurrence of global earthquakes between 1973 to 2011 relative to the longitudinal positions of the sun and moon and compares the result to a simulation of randomly generated earthquakes. 
The data was downloaded from the United States Geological Survey (USGS) website~\cite{USGSEQdata}.
Earthquakes of magnitude 5 and greater were extracted from a date range of January 1, 1973 to July 29, 2011 (the range available when the project was started). 

\begin{figure}[!h]
\begin{center}
\includegraphics[width=4in]{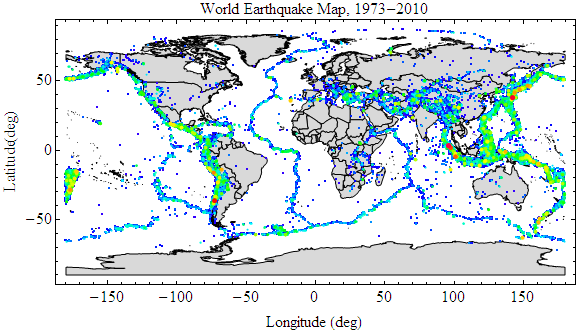}
\end{center}
\caption{
{\bf Earthquakes from January 1, 1973 to July 29, 2011, magnitude 5 and greater.}  The color of the dots denotes magnitude, from blue (magnitude 5) to red (highest magnitude, 9.1).
}
\label{WorldEQs}
\end{figure}

As mentioned earlier, lesser magnitude earthquakes, particularly in the  magnitudes below 4, were not globally recorded as accurately as those of higher magnitudes. 
Therefore the analysis was limited to magnitude 5 and above which encompasses larger earthquakes.
The resulting dataset consisted of 66,724 earthquakes located around the world. Figure~\ref{WorldEQs} shows a visualization of each earthquake on a world map generated with the data.

The following sections describe the method used for establishing the relative difference in position between the moon and sun and each earthquake as well as the histogram method used for the analysis. 

\subsection*{One Hour Binning}\label{Methods_1HrBins}
To evaluate the possibility of the moon influencing earthquakes, it was necessary to calculate the relative difference in position between the moon and each earthquake.
This was done by subtracting the right ascension (RA) coordinates (calculated using Mathematica's ephemerides tables which relied on~\cite{Ephemerides}) for the moon from the longitudinal coordinate of each earthquake, converted to units of time. 
\begin{equation}
d_{rel}=  Longitude_{EQ}(hrs) - RA_{moon}(hrs)
\end{equation}
The result yielded the relative longitudinal difference between the Earth position of each earthquake and the corresponding moon position (latitude offsets were considered less important since the moon moves primarily circumferentially around the Earth). 

\begin{figure}[!h]
\begin{center}
\includegraphics[width=6in]{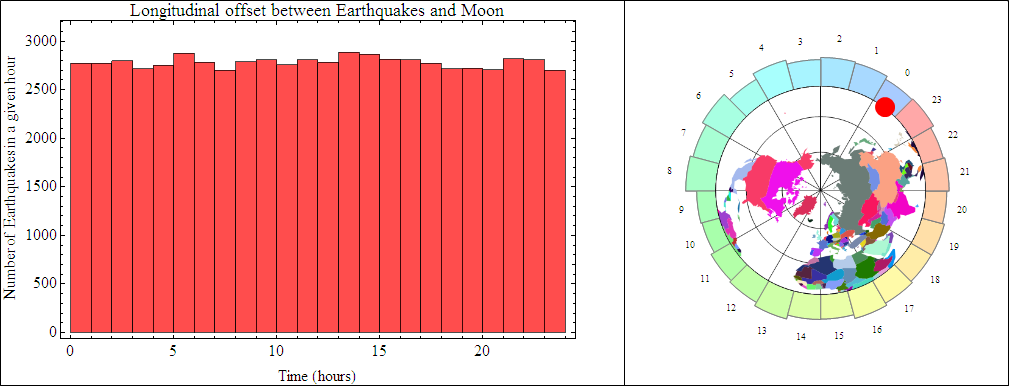}
\end{center}
\caption{
{\bf Histograms of earthquake-moon longitudinal offsets binned by hour.}  The left hand side shows the traditional histogram format of the entire moon-earthquake data. 
The right hand side shows the results referenced to an example earthquake longitude. 
Note that the results are relative to earthquake longitudes in both histograms, instead of referenced to a specific longitude. 
The expected mean number of earthquakes/hour was 2780+/-51.2.
}
\label{RelHistoEQs}
\end{figure}

The longitudinal offsets were then binned according to relative time differences in hours. 
The left hand side of figure~\ref{RelHistoEQs} shows the entire dataset of calculated longitudinal offsets in a traditional histogram format, binned by hour. 
The right hand side of figure~\ref{RelHistoEQs} is an alternate visualization. The 24 wedges represent the 24 hours of possible relative longitudinal distances around the globe from a given earthquake. 
The height of the wedge reflects the number of earthquakes occurring during that offset hour. 
The visualization shows the geometrical relationship between the earthquake, the moon's RA coordinate and the number of earthquakes for each distance.

An examination of the histograms shows slight variations in the number of earthquakes versus the position of the moon.
But does this imply a slight increase in the chance of an earthquake during those moon-earthquake offsets?

In a noiseless dataset, the total number of earthquakes would be divided evenly among the 24 hours of possible longitudinal offsets. 
This would imply 2780 earthquakes per hour of relative offset with a standard deviation of 0 (during the dataset time period, 38.6 years).
However, since the dataset is limited to 66,724 data points, it is necessary to run a simulation that generates the equivalent number of random earthquakes and compare the results to the real-world data to determine if the fluctuations bear any statistical significance.

This simulation was performed by randomly generating longitudinal offsets occurring throughout the day across the same year range as the real-world data.
The distribution along fault lines was implicit as the calculation only used relative distances. 
In that way we don't need to know the positioning of the fault lines. 
As mentioned, this was a source of error since local geography was not taken into account.
In all, 70 sets of 66,724 earthquakes were generated to achieve a stable group of datasets.

\begin{figure}[!h]
\begin{center}
\includegraphics[width=3in]{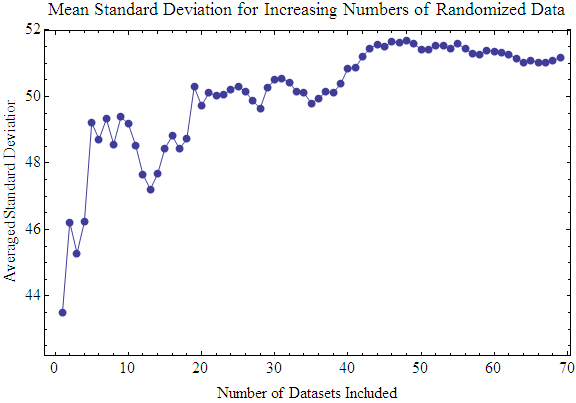}
\end{center}
\caption{
{\bf Stabilization of the standard deviation of the data.} As the number of datasets of randomly generated earthquakes increases, the standard deviation of the height of the averaged bins stabilizes.  
Each dataset has a full 66,724 points.
}
\label{moonEQBGstability}
\end{figure}

The stability was measured by calculating the standard deviation of the hourly bins from the histograms generated from \begin{math}n\end{math} random datasets. 
Equations~\eqref{hist1} and~\eqref{hist2} show how the averaged standard deviation was computed.
As more and more datasets were included, the averaged standard deviation converged.

\begin{equation}
stdev_{bins} = {stdev(\sum_{1}^{n} bin_{hour})}\Bigl\lvert_{each\,hour}
\label{hist1}
\end{equation}
\begin{equation}
stdev_{avg} = \frac{\sum_{1}^{24}{stdev_{bins}}}{n}
\label{hist2}
\end{equation}
Figure~\ref{moonEQBGstability} shows the averaged standard deviation for the averaged hourly bins as more and more random datasets are included.

The resulting simulation predicted a mean of 2780 earthquakes with a standard deviation of  \begin{math}\pm\end{math} 51.2 earthquakes.

\begin{figure}[!h]
\begin{center}
\includegraphics[width=6in]{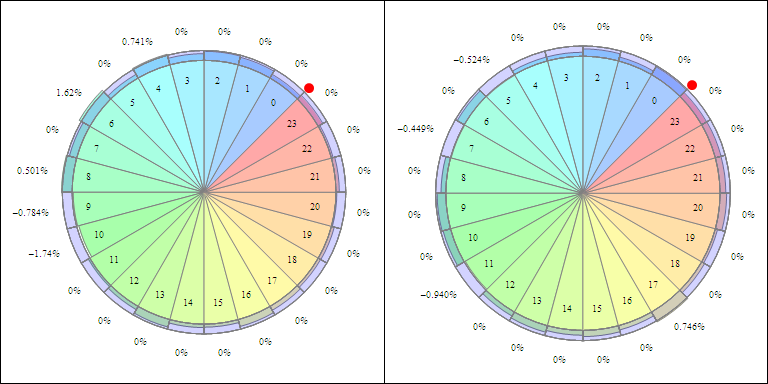}
\end{center}
\caption{
{\bf Relative histograms between earthquake longitude and moon's (left) and sun's (right) longitude (mapped back to Earth).}   
The left shows the same histogram data as figure~\ref{RelHistoEQs} with an overlay of the expected variation (95\% confidence) shown as a gray ring.
The actual increase or decrease is listed as a percentage outside each wedge. 
The inner number shows the relative displacement in hours between the earthquake's and the moon's longitudes.
The expected mean was 2,780 earthquakes +/- 51.2
}
\label{moonEQPlusBG}
\end{figure}

Finally, figure~\ref{moonEQPlusBG} shows the error estimate overlaid on top of the actual measured offset data.
The gray overlapping ring shows \begin{math}\pm1.96\times\end{math}(standard deviation) which implies a 95\% confidence level (the threshold used throughout the paper).
A first examination shows that indeed, some moon-earthquake offsets show a slight increase (and decrease) compared to what would be expected from totally random earthquakes.
A more detailed discussion follows in the Results section of this paper.

\subsection*{Six Hour Binning}\label{Methods_6Hrs}
\begin{figure}[!h]
\begin{center}
\includegraphics[width=2in]{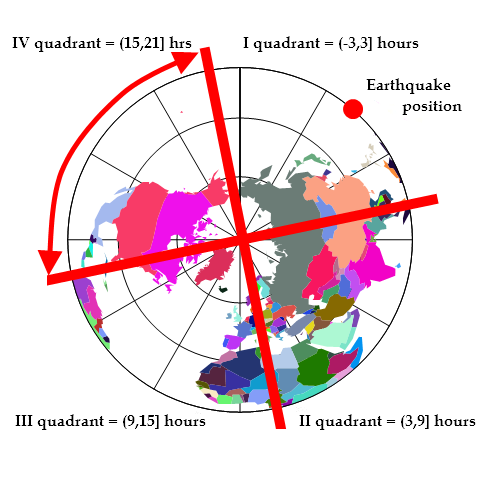}
\end{center}
\caption{
{\bf The Earth divided into 4 quadrants reflecting the 4 phases of the tide referenced to an earthquake's longitude (shown with a red dot).}  The longitude indicated is merely an example. Results are relative to all earthquake data longitudes rather than with respect to an absolute longitude.
}
\label{fig:World4Quadrants}
\end{figure}

In addition to hourly binning, these methods were applied to six-hour longitudinal bins.
Oceanic tides cycle through two high and two low tides lasting six hours over a period of a day.
The high tides correspond to the moon sitting nearly overhead and below the observational longitude, while the low tides correspond to the moon near the horizons which are six hours away, hence the choice of six hour bins.
While there is about a 1 hour lag between the position of the moon and oceanic tides, there is no such lag for land tides~\cite{SAONASAinPopularAstronomy}.
This allows the data to be re-binned into the following six hours increments which capture the entire phase of a tide: 21 hours to 3 hours (which is equivalent to -3 hours to 3 hours), 3 hours to 9 hours, 9 hours to 15 hours and 15 hours to 21 hours (see figure~\ref{fig:World4Quadrants}).

\subsection*{Moon-Sun Position Combinations}
Rather than limiting the analysis to just the moon and sun individually, the relative positions of both the moon and sun in conjunction proved to be important. 
This is the final analytical method used on the data and provided the most interesting set of results. 

\section*{Results}\label{sec:results}

\subsection*{Moon and Sun Considered Separately}\label{hourly_results}
\subsubsection*{Hourly Bins}

\textit{A priori}, it might be guessed that the moon and sun have a low influence on larger magnitude earthquakes.
This proved true.
The statistical significance of hourly binning of earthquakes with respect to the moon's position produced nothing extraordinary. 
The same was true for the sun's relative position. 
Tables~\ref{table:moonZTable} and~\ref{table:sunZTable} at the end of the paper show the number of earthquakes and Z-values for the moon and sun across the full 24 hours of relative longitudinal distance (the Z-value is defined as how many standard deviations the result lies away from the mean).
This data was also presented in figure~\ref{moonEQPlusBG} which shows the moon's results on the left and the sun's results on the right.

The results show that the moon and sun have a tiny effect on larger terrestrial earthquakes.
Over the course of 38.6 years, the moon showed an increase in the likeliness of earthquakes at 4, 6 and 8 hour longitudinal offsets. 
Specifically, the moon-earthquake displacements with the largest associations (at the 95\% confidence level):
\begin{itemize}
\item 4 hours \begin{math} \implies \end{math} 0.741\% greater earthquakes 
\item  6 hours \begin{math} \implies \end{math} 1.62\% greater earthquakes
\item  8 hours \begin{math} \implies \end{math} 0.501\% greater earthquakes
\end{itemize}
Dividing by the 38.6 year duration, this amounted to a combined total of 2.2 extra earthquakes per year, magnitude 5 and greater during those longitudinal offsets.

Similar to the moon, the sun showed an effect on terrestrial earthquakes, although the association was even weaker.
Earthquakes were more likely than expected when the sun was 17 hours behind a geographical longitude (Z=2.38, 0.746\% increase, 95\% confidence).

On the other hand, certain longitudinal offsets for the sun and moon reduced the chance of an earthquake.
In total, the moon was associated with a reduction of 1.8 earthquakes per year during lower risk offsets (9th and 10th hours).
The sun was associated with 1.3 fewer earthquakes per year during the longitudinal offsets of 5, 7 and 11 hours.

\subsubsection*{Six Hour Bins}
The associations were non-existent for six-hour bins, defined according to the tidal cycles (figure~\ref{fig:World4Quadrants}). 
Tables~\ref{table:moonQuadZTable} and~\ref{table:sunQuadZTable} show the associations of the sun and moon with the relative positions of earthquakes.
In retrospect, the averaging out of signal in the six hour bins should have been obvious due to the haphazard nature of the increases and decreases in the hourly bins.

Aside from the minimal influence, one issue in particular stood out in this analysis: there was no clear cycle present.
Reference~\cite{BigAnalysis} showed a daily cycle similar to the water tide, although the pattern was for shallow, low-magnitude earthquakes which have different origins than high magnitude earthquakes.

The next section examines combined sun-moon positions. 
Surprisingly, this approach resulted in a significant changes in earthquakes for certain moon-sun pairings that were cyclical in nature.

\begin{table}[!h]
\centering
\caption{\bf{The Z-value associated with moon-earthquake relative distances with 6-hour binning}}
\begin{tabular}{|c|c|c|c|}
\hline
Moon (quad) &  EQs & Z-value & \% extra EQs\\
\hline
I & 16676 &  -0.0441 & 0\\
II & 16612 & -0.552 & 0\\
III & 16901 &  1.76 & 0\\
IV & 16535 & -1.17 & 0\\
\hline
\end{tabular}
\begin{flushleft}
The mean and standard deviation of the equivalent size dataset of randomly generated earthquakes was 16,681 \begin{math}\pm\end{math} 125.
\end{flushleft}
\label{table:moonQuadZTable}
\end{table}

\begin{table}[!h]
\centering
\caption{\bf{The Z-value associated with sun-earthquake relative distances with 6-hour binning}}
\begin{tabular}{|c|c|c|c|}
\hline
Sun (quad) &  EQs & Z-value & \% extra EQs\\
\hline
I & 16818 &  1.10 & 0\\
II & 16463 & -1.74 & 0\\
III & 16692 &  0.088 & 0\\
IV & 16751 & 0.056 & 0\\
\hline
\end{tabular}
\begin{flushleft}
The mean and standard deviation of the equivalent size dataset of randomly generated earthquakes was 16,681 \begin{math}\pm\end{math} 125. 
The sun's results produced low statistical fluctuations.
\end{flushleft}
\label{table:sunQuadZTable}
\end{table}

\subsection*{Combined Moon-Sun Positions}\label{binningtogether}
\subsubsection*{Six Hour Binning}\label{binningtogether6}
While looking at the moon and sun separately revealed little, looking at moon-sun combinations did result in a significant influence. 
Unlike the expected mean number of earthquakes for the moon and sun considered separately, the combined moon-sun ephemerides caused the mean earthquakes to vary by each bin.
The effect was small  (\textless 1\%) since the bin widths were relatively large, but this was still taken into account when calculating the mean expected values for each of the six-hour bins.

Whenever the moon and sun paired together in the horizon, there was a significant decrease in earthquakes. 
Table~\ref{table:moonsunOppositesTable} lead to the following conclusions for the (moon, sun) quadrants with a 95\% confidence level:
\begin{itemize}
\item (II,II) \begin{math} \implies \end{math} 4.57\% fewer earthquakes
\item  (IV,IV) \begin{math} \implies \end{math} 2.31\% fewer earthquakes
\end{itemize}
One physical interpretation is that when the moon and sun are paired together on the horizons they act to horizontally pull the Earth's crust together, helping to seal fault lines.
This force behaves differently than when the moon and sun are overhead and pull both sides of a fault upward rather than sideways.

Conversely, the statistical increases in earthquakes were shallow and occurred when the moon and sun were not paired in the same quadrant:
\begin{itemize}
\item (II,IV) \begin{math} \implies \end{math} 1.08\% increase in earthquakes
\item (III,I) \begin{math} \implies \end{math} 1.21\% increase in earthquakes
\item (III,IV) \begin{math} \implies \end{math} 0.345\% increase in earthquakes
\end{itemize}

It seems intuitive that when the moon and sun are pulling on opposite sides of a fault they separate the two halves just enough to trigger an earthquake.

What about the other permutations? 
Table~\ref{table:moonsunOppositesTable} shows the measured earthquakes versus expected earthquakes for all possible moon and sun quadrant pairings.

\begin{table}[!h]
\centering
\caption{\bf{The Z-value associated with all possible moon and sun quadrant permutations}}
\begin{tabular}{|c|c|c|c|c|c|}
\hline
Moon & Sun &  EQs & expected EQs & Z-value & \% extra EQs\\
\hline
I & I & 4025 & 4136 & -1.77 & 0\\
I & II & 4295 & 4185 & 1.74 & 0\\
I & III & 4190 & 4156 & 0.539 & 0\\
I & IV & 4166 & 4197 & -0.489 & 0\\
II & I & 4196 & 4191 & 0.0848 & 0\\
II & II & 3848 & 4147 & -4.76 & -4.57\\
II & III & 4244 & 4195 & 0.781 & 0\\
II & IV & 4324 & 4155 & 2.7 & 1.08\\
III & I & 4316 & 4141 & 2.8 & 1.21\\
III & II & 4133 & 4194 & -0.975 & 0\\
III & III & 4115 & 4133 & -0.286 & 0\\
III & IV & 4337 & 4199 & 2.2 & 0.345\\
IV & I & 428I & 4202 & 1.26 & 0\\
IV & II & 4187 & 4154 & 0.534 & 0\\
IV & III & 4143 & 4203 & -0.95 & 0\\
IV & IV & 3924 & 4136 & -3.41 & -2.31\\
\hline
\end{tabular}
\begin{flushleft}Based on randomly generated datasets of simulated earthquakes, the standard deviation was \begin{math}\pm\end{math}62.7.
The number of expected earthquakes varied by bin and was therefore listed in a separate column in the table.
\end{flushleft}
\label{table:moonsunOppositesTable}
\end{table}

\subsubsection*{One Hour Binning}\label{binningtogether1}

\begin{figure}[!h]
\begin{center}
\includegraphics[width=6in]{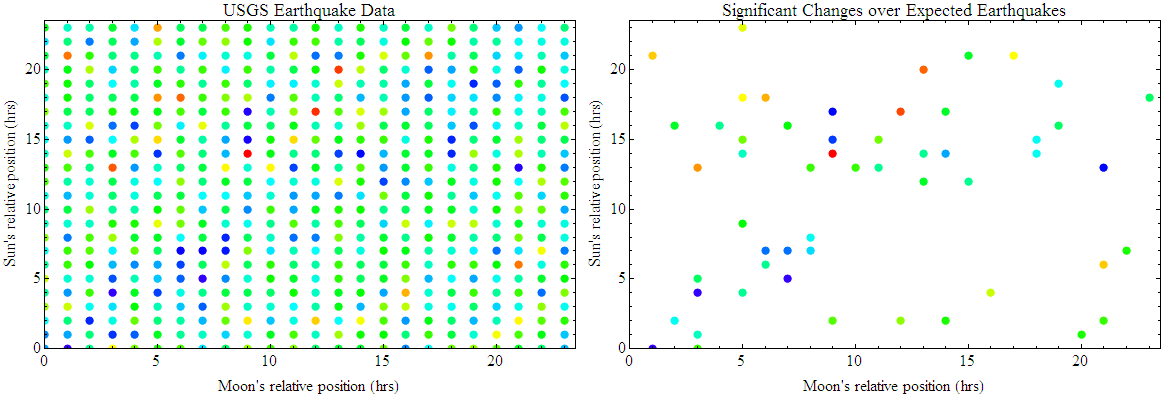}
\end{center}
\caption{
{\bf Earthquakes for various moon/sun offset positions.}
A plot of the moon/sun positions vs. the number of earthquakes at those positions. The redder the dot, the larger the number of earthquakes, the bluer the dot, the fewer.
The plot on the right shows the same data with only the 95\% confidence results. The positions which lay within expected variation were removed.
}
\label{EarthquakesWRTMoonSun3D}
\end{figure}

The final step was to calculate the hourly combined position results. 
The mean expected earthquakes varied more with hourly binning, amounting to a 4.4\% effect as opposed to \textless 1\% with six-hour bins.
Taking this into account, a large table of 576 bins resulted, of which most combinations produced null or negligible increases.
However, 32 positions (about 5.6\% of the possible permutations) were statistically significant. 
An initial plot of the moon's and sun's combined positions versus the increase in expected earthquakes showed the apparently random data (left, figure~\ref{EarthquakesWRTMoonSun3D}).
The right hand side of figure~\ref{EarthquakesWRTMoonSun3D} showed the results if only the statistically significant results were plotted.
Table~\ref{table:sunmoonZTable} at the end of the paper lists the values for the non-negligible results (using \begin{math}Z>2.5\end{math} so the chart could fit on one page).

A moment should be taken to examine the true significance of the results in the table and figures.
Some of the results show very sharp increases and decreases over expected, right next to an hourly bin combination that had no significant difference.
For instance, (moon,sun)=(3,4) hrs offset resulted in 13\% fewer earthquakes compared to the mean expected number.
However, (moon,sun)=(4,4) hrs offset was within the expected range of variation.
Clearly, even though the noise instrinsic to the size of the dataset was taken into account, another source of noise must still be present.
As mentioned at the start of the paper, the method of looking only at the relative distance between an earthquake and the celestial positions of the moon and sun ignored localized geological properties that affect the heights of land tides.
It is likely that this influence added an extra significant source of noise, on top of the noise intrinsic to the dataset.

A new approach was thus proposed: treat the statistically significant points  listed in Table~\ref{table:sunmoonZTable} as outliers on top of a noisy dataset.
Examine the outliers for a periodic trend.
Periodicity in the result adds validity since the system is cyclical in nature.
Then, based on the pattern identified, select the broadest possible bins in order to average out the influence of the extra noise while still preserving the earthquake "signal". 

\subsubsection*{Outlier Analysis}\label{OutlierAnalysis}
\begin{figure}[!h]
\begin{center}
\includegraphics[width=6in]{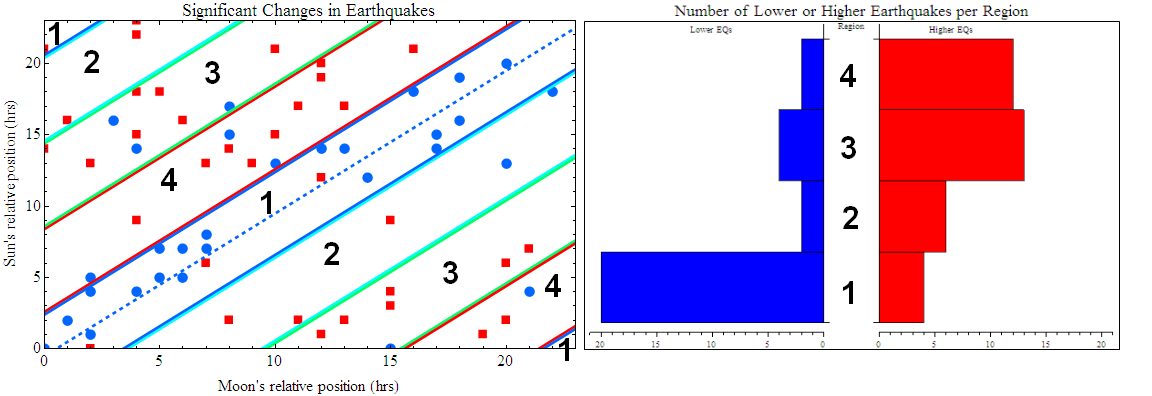}
\end{center}
\caption{
{\bf Significant changes in earthquakes for various moon/sun offset longitudes.}
On the left, a plot shows the moon/sun relative longitudes. 
The blue circles show where there were fewer than expected earthquakes and the red squares show where there were higher than expected earthquakes.  
Because the day cycles every 24 hours, certain regions are extensions of each other and have been labeled to show this relationship.
The diagram on the right shows a histogram of the lower than expected earthquakes (blue) and higher than expected earthquakes (red) in the labeled regions.
Region 1 (when the sun was +2.5 or -3.5 hours of the moon, independent of longitude) showed reduced earthquakes. 
Regions 3 and 4 showed increases in high earthquake moon-sun pairings.
}
\label{significantEQs}
\end{figure}

Ignoring the specific value of the outliers and plotting the fewer-than-expected earthquake bins as blue circles and the higher-than-expected earthquake bins as red squares, a cyclical pattern emerged.
Figure~\ref{significantEQs} shows that whenever the sun was +2.5 or -3.5 hours away from the moon, \textit{irrespective of longitude}, there was a trend toward a reduction of earthquakes.

The axes in figure~\ref{significantEQs} were labeled such that:
\begin{equation}
x =  Longitude_{EQ}(hrs) - RA_{moon}(hrs)
\end{equation}
\begin{equation}
y = Longitude_{EQ}(hrs) - RA_{sun}(hrs)
\end{equation}

Thresholds were selected to identify the region of reduced earthquakes as three hours wide around the central, dotted line, \begin{math}y=x-1/2\end{math}. 
This bin definition captured the most blue points and a minimum of red points.
Labeled regions in the figure identify the bin definitions.
Because the days cycle every 24 hours, some regions are extensions of others, reflected in the labeling.

\begin{itemize}
\item Region 1 accounted for 67\% of the fewer-than-expected earthquake moon-sun positions, and only 11\% of the higher-than-expected earthquake combinations fell in between these thresholds.
\item Region 2 had only 13\% of the lowered combinations and 34\% of the higher combinations.
\item Region 3 had 13\% lowered combinations and 37\% higher combinations.
\item Region 4 had 7\% lowered combinations and 17\% higher combinations.
\end{itemize}

Do the outliers predict real earthquake trends? 
The question is answered by summing the earthquakes data in the four regions and again comparing to the 70 randomly generated datasets.
These 4 regions encompassed about 16,700 points each as opposed to the 116 points for the hourly bins which reduces random noise fluctuations by a factor of six.
The mean expected earthquakes for each bin were recalculated for the four regions.
Table~\ref{table:regionsTable} shows the results.

\begin{table}[!h]
\centering
\caption{\bf{The Z-value associated with the four regions depicted in Figure~\ref{significantEQs}}}
\begin{tabular}{|c|c|c|c|c|}
\hline
Region & EQs & expected EQs & Z-value & \% extra EQs\\
\hline
1 & 15714 & 16509 & -6.36 & -3.33\\
2 & 16776 & 16832 & -0.445 & 0 \\
3 & 17169 & 16554 & 4.92 & 2.23\\
4 & 17065 & 16829 & 1.89 & 0\\
\hline
\end{tabular}
\begin{flushleft}
Based on randomly generated datasets of simulated earthquakes, the standard deviation of earthquakes is +/- 125.
\end{flushleft}
\label{table:regionsTable}
\end{table}

Indeed, the abundance of outlier lower-than-expected earthquakes correctly identified a region of generally reduced earthquakes.
A reduction of 3.33\% earthquakes (95\% confidence) implied about 14.3 fewer large magnitude earthquakes per year when the sun was +2.5 or -3.5  hours away from the moon, independent of longitude. 

Conversely, Region 3 (the sun was 8.5 to 14.5 hours behind the moon) showed significantly higher than expected earthquakes, 2.23\% at the 95\% confidence level.
This amounted to 9.6 extra earthquakes per year.

\begin{figure}[!h]
\begin{center}
\includegraphics[width=6in]{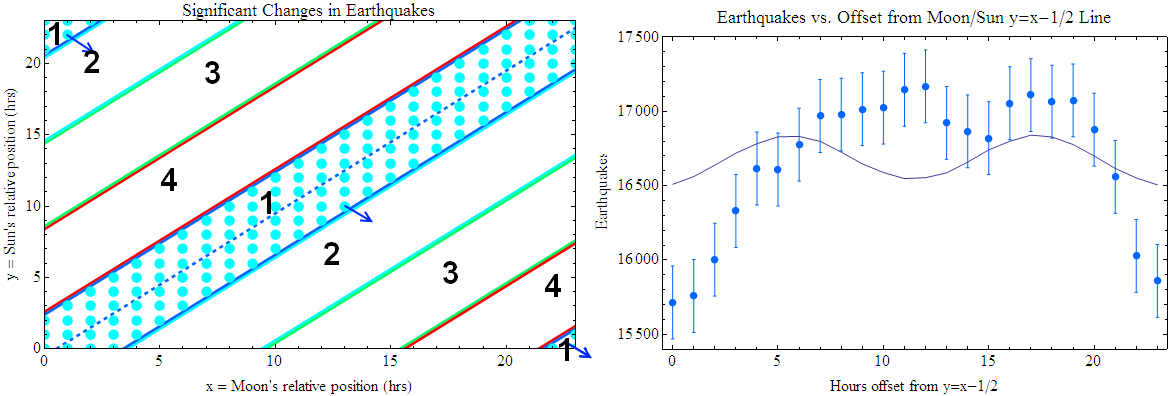}
\end{center}
\caption{
{\bf Cyclical pattern from summing earthquakes by stepping through offsets to y=x-1/2.}
Stepping through hour-by-hour earthquakes binned +/-3 hours around the line y=x-1/2. For example, Region 1 corresponds to the 0 hours offset from y=x-1/2 (+/-3 hours) and Region 3 corresponds to the 11 hours offset (+/-3 hours).
The right hand side shows the raw summed earthquakes plotted with estimated error bars.
The sinusoidal line shows the estimated mean background for each of the positions, a number which varied due to the ephemerides of the combined moon-sun positions.
}
\label{fig:stepEQs}
\end{figure}

Starting in Region 1 and stepping the 6-hour bin through the moon-sun earthquake space an hour at a time revealed a cyclical pattern, seen in figure~\ref{fig:stepEQs}. 
The minimum earthquake total occured for Region 1 and the maximum aligned with Region 3. 
The sinusoidal line shows the estimated mean background for each of the steps, a number which varied due to the ephemerides of the combined moon-sun positions.

\begin{figure}[!h]
\begin{center}
\includegraphics[width=3in]{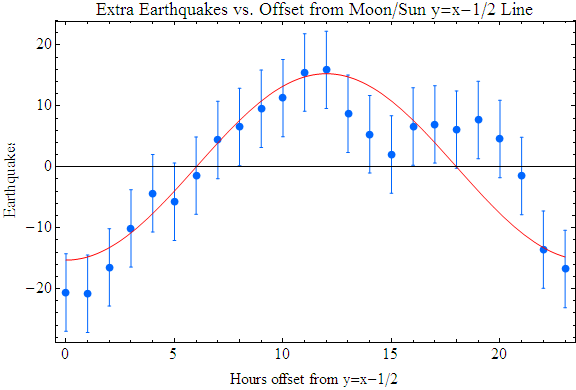}
\end{center}
\caption{
{\bf Cyclical pattern from summing earthquakes by stepping through offsets to y=x-1/2.}
Stepping through hour-by-hour earthquakes binned +/-3 hours around the line y=x-1/2. 
For example, the 0 hours offset corresponds to Region 1 and the 12 hour offset corresponds to Region 3.
}
\label{fig:Final_Plot}
\end{figure}

Plotting the difference between the mean expected earthquakes and the measured earthquakes produced figure~\ref{fig:Final_Plot}. 
The data for the plot was recorded in Table~\ref{table:regionsTable}. 
Fitting an equation to the data gave:
\begin{equation}
EQ_{extra}=-15.3 Cos[\frac{2Pi}{24}*offset_{y=x-1/2}]
\end{equation}
This equation gives a simple relationship between the measured number of earthquakes over (or less than) the mean expected number of earthquakes which depends strictly on the relative distance between the moon and sun.
It is also notable that these conclusions agree with the more narrow conclusions of 6-hour combination binning relative to a specific longitude (Table~\ref{table:moonsunOppositesTable}).

\section*{Discussion}\label{Discussion}
\noindent 
The earthquake data, despite coming from a large, accurate record of events, was still noisy.
The assumption in the paper was that the method of looking at relative offsets did not account for local properties that influenced land tide heights and, in turn, influenced the likeliness of an earthquake.
The only way that local geography could possibly influence the results would be if specific faults or fault types were over-represented in the outlier moon-sun combinations.
Then the local geography would exert an influence.
If this was not the case, but the fault lines and types were randomly scattered in the moon-sun earthquake space then local geography could not be a noise source and something else must be influencing the results.
A future step to research would be to examine the specific fault and fault type distributions in the data in more detail to verify this statement. 

Despite this, by examining outliers, a clear association of the positions of the moon and sun and earthquakes was identified. 
Surprisingly, the clearest trend was for a reduction of earthquakes when the moon and sun were close together in the sky, independent of the longitude of an earthquake.
When the moon and sun were on opposite sides of the Earth (Region 3) there was a marked increase in large magnitude earthquakes. 

The fact that the location of the earthquake was irrelevant may be related to the depth of the earthquakes.
Deeper fault lines, closer to the Earth's core, may be less influenced by the circumferential effects of the moon and sun.

It is interesting to note that although the association was based on a statistical analysis of a large dataset, the pattern holds for three of the four highest-magnitude earthquakes in the study, including:
\begin{itemize}
\item December 2004 Northern Sumatra earthquake (magnitude 9.1, relative sun RA=0.227, relative moon RA=12.5)
\item March 2005 Indonesia earthquake (magnitude 8.6, relative sun RA=9.17, relative moon RA=18.8) and
\item February 2010 Chile earthquake (magnitude 8.8, relative sun RA=4.09, relative moon RA=15.0)
\end{itemize}

As always, correlation does not necessarily imply causation.
One alternate source for the results of the study could be an estimate for a correction to the moon/sun ephemerides. 
The appearance of an increase or decrease in earthquakes could imply an error in the calculated moon/sun positions at the time of the earthquakes.
This paper used Mathematica to generate the RA coordinates which which employed the current standard, VSOP87 (Variations Séculaires des Orbites Planétaires).
The precision of this standard is stated as 1" per 4000 years around the year 2000 (see reference~\cite{Ephemerides}), making the possibility of the bias unlikely.

\section*{Acknowledgments}
The author expresses thanks to the USGS for posting their earthquake data online and Mathematica for making this project possible in a finite amount of time.

\bibliography{EarthquakesMoonSun}


\section*{Tables}

\begin{table}[!h]
\centering
\caption{\bf{The Z-value associated with moon-earthquake relative distances}}
\begin{tabular}{|c|c|c|c|}
\hline
Moon (hrs) &  EQs & Z-value & \% extra EQs\\
\hline
0 & 2767 & -0.257 & 0\\
1 & 2840 & 1.17 & 0\\
2 & 2863 & 1.62 & 0\\
3 & 2816 & 0.7 & 0\\
4 & 2902 & 2.38 & 0.741\\
5 & 2765 & -0.296 & 0\\
6 & 2928 & 2.89 & 1.62\\
7 & 2834 & 1.05 & 0\\
8 & 2895 & 2.24 & 0.501\\
9 & 2659 & -2.37 & -0.784\\
10 & 2634 & -2.86 & -1.74\\
11 & 2715 & -1.27 & 0\\
12 & 2755 & -0.492 & 0\\
13 & 2836 & 1.09 & 0\\
14 & 2751 & -0.57 & 0\\
15 & 2742 & -0.746 & 0\\
16 & 2825 & 0.876 & 0\\
17 & 2772 & -0.16 & 0\\
18 & 2734 & -0.902 & 0\\
19 & 2740 & -0.785 & 0\\
20 & 2698 & -1.61 & 0\\
21 & 2748 & -0.628 & 0\\
22 & 2716 & -1.25 & 0\\
23 & 2789 & 0.173 & 0\\
\hline
\end{tabular}
\begin{flushleft}
A positive Z-value reflects more earthquakes than expected while a negative Z-value implies less earthquakes than expected. 
A simulation was used to calculate the mean number of earthquakes and the mean standard deviation of the equivalent size dataset of randomly generated earthquakes. 
This produced the values 2780 \begin{math}\pm\end{math} 51.2. 
As can be seen in the table, the results had mostly no or low statistical significance.
\end{flushleft}
\label{table:moonZTable}
\end{table}

\begin{table}[!h]
\centering
\caption{\bf{The Z-value associated with sun-earthquake relative distances}}
\begin{tabular}{|c|c|c|c|}
\hline
Sun (hrs) &  EQs & Z-value & \% extra EQs\\
\hline
0 & 2856 & 1.48 & 0\\
1 & 2758 & -0.434 & 0\\
2 & 2823 & 0.838 & 0\\
3 & 2757 & -0.453 & 0\\
4 & 2759 & -0.414 & 0\\
5 & 2666 & -2.23 & -0.524\\
6 & 2839 & 1.15 & 0\\
7 & 2668 & -2.19 & -0.449\\
8 & 2774 & -0.121 & 0\\
9 & 2850 & 1.37 & 0\\
10 & 2830 & 0.975 & 0\\
11 & 2655 & -2.45 & -0.94\\
12 & 2771 & -0.179 & 0\\
13 & 2811 & 0.603 & 0\\
14 & 2775 & -0.101 & 0\\
15 & 2801 & 0.408 & 0\\
16 & 2766 & -0.277 & 0\\
17 & 2902 & 2.38 & 0.746\\
18 & 2747 & -0.649 & 0\\
19 & 2725 & -1.08 & 0\\
20 & 2810 & 0.584 & 0\\
21 & 2793 & 0.251 & 0\\
22 & 2796 & 0.31 & 0\\
23 & 2792 & 0.231 & 0\\
\hline
\end{tabular}
\begin{flushleft}
A simulation showed the mean number of earthquakes during the time-period to be 2780 \begin{math}\pm\end{math} 51.2. 
The sun produced even lower statistically significant results than the moon.
\end{flushleft}
\label{table:sunZTable}
\end{table}

\begin{table}[!h]
\centering
\caption{\bf{The non-zero values for earthquakes exceeding (or less than) the expected variation in earthquakes at the 95\% confidence level. The moon-sun-earthquake relative distances.}}
\begin{tabular}{|c|c|c|c|c|}
\hline
Moon (hrs) & Sun (hrs) &  EQs & Z-value & \% extra EQs\\
\hline
1 & 0 & 85 & -2.85 & -10.9\\
1 & 21 & 150 & 3.17 & 8.48\\
3 & 0 & 141 & 2.52 & 4.16\\
3 & 4 & 85 & -3.02 & -13.1\\
3 & 13 & 152 & 3.47 & 10.4\\
5 & 4 & 93 & -2.3 & -3.78\\
5 & 14 & 92 & -2.22 & -2.93\\
5 & 15 & 143 & 2.76 & 5.86\\
5 & 18 & 148 & 3.2 & 8.79\\
5 & 23 & 147 & 2.81 & 6.06\\
6 & 7 & 88 & -2.25 & -3.42\\
6 & 18 & 151 & 3.55 & 11.1\\
7 & 5 & 85 & -2.78 & -10.2\\
7 & 7 & 88 & -2.46 & -5.93\\
8 & 7 & 90 & -2.34 & -4.45\\
9 & 14 & 158 & 3.79 & 12.2\\
9 & 15 & 87 & -2.9 & -11.4\\
9 & 17 & 86 & -2.91 & -11.6\\
10 & 13 & 141 & 2.55 & 4.39\\
11 & 15 & 143 & 2.55 & 4.35\\
12 & 2 & 144 & 2.77 & 5.91\\
12 & 17 & 155 & 3.58 & 11.\\
13 & 20 & 154 & 3.33 & 9.31\\
14 & 2 & 140 & 2.38 & 3.17\\
14 & 14 & 89 & -2.55 & -6.91\\
16 & 4 & 146 & 3.05 & 7.84\\
17 & 21 & 148 & 2.92 & 6.84\\
18 & 14 & 91 & -2.47 & -5.94\\
18 & 15 & 91 & -2.42 & -5.36\\
21 & 6 & 150 & 3.22 & 8.81\\
21 & 13 & 86 & -2.91 & -11.6\\
23 & 18 & 94 & -2.26 & -3.3\\
\hline
\end{tabular}
\begin{flushleft}
The mean of the equivalent size dataset of randomly generated earthquakes varied by position according to the moon and sun ephemerides (calculated via Mathematica). The means ranged from 111 to 120 earthquakes during the dataset time span (38.6 years). The standard deviation was \begin{math}\pm\end{math} 10.8. 
The large number of statistically significant results is attributed to noise introduced by the method of analysis (relying on the position of the sun and moon to gauge the effect of land-tides on earthquake fault lines).
\end{flushleft}
\label{table:sunmoonZTable}
\end{table}

\begin{table}[!h]
\centering
\caption{\bf{The Z-value associated with the y=x-1/2 region stepped through all 24 hours, as depicted in Figure~\ref{fig:stepEQs}}}
\begin{tabular}{|c|c|c|c|c|c|}
\hline
Region & EQs & expected EQs & Z-value & \% extra EQs & extra EQ/yr\\
\hline
0 & 15714 & 16509 & -6.36 & -3.33 & -14.3\\
1 & 15758 & 16561 & -6.43 & -3.37 & -14.5\\
2 & 16002 & 16638 & -5.09 & -2.35 & -10.1\\
3 & 16330 & 16719 & -3.11 & -0.863 & -3.74\\
4 & 16616 & 16783 & -1.34 & 0 & 0\\
5 & 16609 & 16829 & -1.76 & 0 & 0\\
6 & 16776 & 16832 & -0.445 & 0 & 0\\
7 & 16969 & 16798 & 1.36 & 0 & 0\\
8 & 16979 & 16726 & 2.02 & 0.0477 & 0.207\\
9 & 17015 & 16647 & 2.95 & 0.74 & 3.19\\
10 & 17025 & 16589 & 3.49 & 1.15 & 4.94\\
11 & 17146 & 16548 & 4.78 & 2.13 & 9.15\\
12 & 17169 & 16554 & 4.92 & 2.23 & 9.59\\
13 & 16923 & 16587 & 2.69 & 0.55 & 2.37\\
14 & 16866 & 16659 & 1.65 & 0 & 0\\
15 & 16820 & 16741 & 0.63 & 0 & 0\\
16 & 17055 & 16800 & 2.04 & 0.0622 & 0.271\\
17 & 17110 & 16841 & 2.15 & 0.144 & 0.629\\
18 & 17065 & 16829 & 1.89 & 0 & 0\\
19 & 17074 & 16777 & 2.37 & 0.308 & 1.34\\
20 & 16877 & 16700 & 1.41 & 0 & 0\\
21 & 16559 & 16616 & -0.46 & 0 & 0\\
22 & 16028 & 16552 & -4.19 & -1.69 & -7.23\\
23 & 15859 & 16506 & -5.17 & -2.43 & -10.4\\
\hline
\end{tabular}
\begin{flushleft}
Based on randomly generated datasets of simulated earthquakes, the standard deviation of earthquakes is +/- 125.
\end{flushleft}
\label{table:regionsTable}
\end{table}

\end{document}